\documentclass[journal]{IEEEtran}
\usepackage{enumitem}
\usepackage{amssymb}
\usepackage{graphicx}
\usepackage{hyperref}
\newcommand{\Var}{\mathbb{V}\text{ar}}
\newcommand{\Cov}{\mathbb{C}\text{ov}}
\usepackage{cancel}
\usepackage{comment}
\usepackage{xcolor}

\usepackage{tikz}
\usetikzlibrary{positioning}
\usetikzlibrary{shapes,arrows}

\ifCLASSOPTIONcompsoc
  \usepackage[nocompress]{cite}
\else
  \usepackage{cite}
\fi
\ifCLASSINFOpdf
\else
\fi

\usepackage{amsmath}

\usepackage{stfloats}
\ifCLASSOPTIONcaptionsoff
 \let\MYoriglatexcaption\caption
 \renewcommand{\caption}[2][\relax]{\MYoriglatexcaption[#2]{#2}}
\fi

\begin{document}
%
\title{Accelerography: Feasibility of Gesture Typing using Accelerometer}

\author{Arindam Roy Chowdhury, 
        Abhinandan Dalal 
        and Shubhajit Sen


    }

\maketitle


%

\textbf{\textit{Abstract} - In this paper, we aim to look into the feasibility of constructing alphabets using gestures. The main idea is to construct gestures, that are easy to remember, not cumbersome to reproduce and easily identifiable. We construct gestures for the entire English alphabet and provide an algorithm to identify the gestures, even when they are constructed continuously. We tackle the problem statistically, taking into account the problem of randomness in the hand movement gestures of users, and achieve an average accuracy of 97.33\% with the entire English alphabet.}

\vspace{5Pt}


\vspace{35Pt}

\IEEEraisesectionheading{\section{Introduction}\label{sec:introduction}}

\IEEEPARstart{I}n this era, smartphones have become an integral part of our lives. They are able to handle, if not solve, many of our daily needs.They have brought in the idea of gesture recognition to make lives simpler.\par
In this paper, we have proposed an approach to use a 3 axis accelerometer in a smartphone to create gestures for the entire English Alphabet. The relevant literature has often discussed the idea of gesture typing, but mainly on a very small alphabet system or system of digits. For instance, \cite{digit} discusses the idea of a motion based pen without a special writing surface, which uses the motion of the pen, angular velocity through accelerometers and gyroscopes to recognize digits. \cite{NN} Proposes an accelerometer based hand gesture recognition algorithm, which uses neural networks. But, neural networks are computationally expensive and require large number of training examples. Further approaches to the problem can be found in \cite{timeseries}, \cite{HMM}.\par
We must keep in mind the statistical nature of the problem. One may define the gestures arbitrarily and ask the user to replicate it. But, hand movements cannot be controlled to the pin-point specification, and hence gestures should be well separated so that it can accommodate the idea of randomness of the hands and still give correct outputs. Thus we need to handle the tradeoff between the intuition of the gestures for its simplicity, as well as its distinguishability from other gestures to identify letters correctly. .Moreover, we emphasize on the use of statistical methodologies over usages of non-interpretable methods (like neural networks) because the data generating process is inherently human, and the focus on essentially statistical methods translates to easier reflection on the concerns from the user parts, i.e., it controls the user error by making him realize what makes the algorithm make errors. Hence we bring in a simple statistical approach, which has an easy to implement algorithm with very good success rates. This approach to the problem, as per our knowledge, has not been discussed in the literature. 

\parskip 1pt

\section{Challenges}

 We aim to devise an approach in which people would be able to type without touching the mobile screen. First, we enlist the challenges we faced to tackle the entire problem. 

\begin{itemize}

    \item We must make the gestures intuitive and easy to remember.
    
    \item We must capture the randomness of people's hands movements. 
    
    \item We define two kinds of error, which we need to control:
    
      \begin{itemize}
          \item \textbf{user error}: This  is caused when the user inputs incorrect gestures for the letters. This may occur if the gestures are confusing. It is not entirely statistical in nature and cannot be given bounds mathematically in absence of data from surveys.
          \item \textbf{classification error}: This occurs when the user's input is miss-classified as a different letter, which the user did not intend to achieve.
      \end{itemize}

    \item Different persons have different speeds of making their gestures. Also, someone may make the gestures using more space, while someone require smaller space. Thus we need to keep in mind the speed and size considerations for our gestures.
    
\end{itemize} 

\section{Heading to the Solution: The Gravity Ink Approach}

\subsection{The Letters}
 Table ~\ref{LetterGI} provides the summarized version of the letters and their corresponding resemblances to make it easier for the user to identify.\par 

\begin{table}[ht]
    \centering
    \begin{tabular}{|c|c|c|c|}
    \hline 
        Letter & Resemblance & Representation & Rel. Freq\\
        \hline \hline
        A & $\wedge$ & U & 8.167\% \\ 
        B & $3$ & RR & 1.492\% \\
        C & C & L & 2.782\% \\ 
        D & D & R & 4.253\% \\
        E & $\varepsilon$ & LL & 12.702\% \\
        F & F & LU & 2.228\% \\ 
        G & $\Gamma$ & UL & 2.015\% \\
        H & H & RL & 6.094\% \\
        I & $=$ & UD & 6.966\% \\
        J & $\_|$ & RD & 0.153\% \\
        K & $\kappa$ & LUD & 0.772\% \\
        L & L & LD & 4.025\% \\
        M & $\wedge\wedge$ & UU & 2.406\% \\
        N & N & RUL & 6.749\% \\
        O & O & ULDR & 7.507\% \\
        P & P & DR & 1.929\% \\
        Q & Q & DRULD & 0.095\% \\
        R & 13 & LRR & 5.987\% \\
        S & S & ULD & 6.327\% \\
        T & t & DLR & 9.056\% \\
        U & U & LDR & 2.758\% \\
        V & V & D & 0.978\% \\
        W & VV & DD & 2.360\% \\
        X & + & UDLR & 0.150\% \\
        Y & Y & LRD & 1.974\% \\
        Z & Z & URLD & 0.074\% \\
        \hline 
    \end{tabular} 
    \caption{Letters and their Representative Gestures: Gestures based on Dextral People's General Hand Movements.  However, defining the complement of Up as Down and the complement of Left as Right, a similar kind of gestures shall work for sinistral people.
}
    \label{LetterGI}
\end{table}

First let us provide the intuition of this approach, and justify its nomenclature. We use the orientation of the phone as the major backbone of this procedure.  Suppose your phone is fixed at its center of gravity. Imagine that there is ink lying on the edges of the phone and there is a paper lying below your phone, and all you are allowed to do is make ink impressions with your phone on the paper. Thus you can make only rectilinear movements due to the fulcrum being at the center of gravity. We shall use the order of the impressions to form the letters. Thus a left movement followed by a right movement is different from a right movement followed by a left movement. \par

\begin{figure}
    \centering
    \includegraphics[height= 0.5\textwidth, width=0.45\textwidth,clip, keepaspectratio]{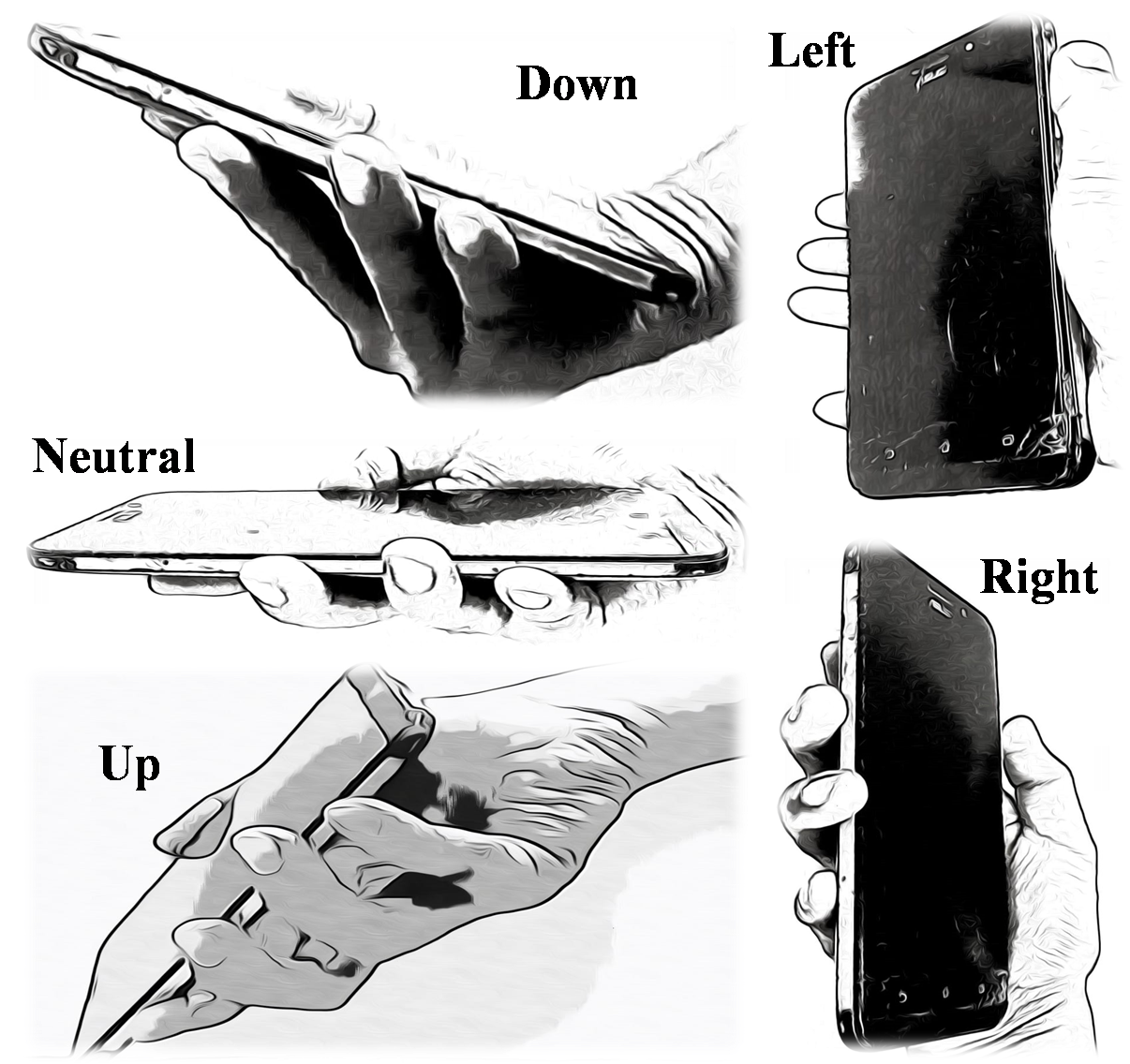}
    \caption{Phone movements showing all the defined user moves}
    \label{orient}
\end{figure}

Hence, for each movement, we have four degrees of freedom,\textbf{ Left: L, Up: U, Right: R, Down: D} [\ref{orient}]. This idea of using basic gestures has also been used in\cite{NN}. Now, we do not want to make our gestures very long, but, in order to make the gestures intuitive and memorable, we have to allow a little longer gestures. This helps reduce the user error. \par 
As the user error gets reduced over time, owing to muscle memory and adaptivity of humans, we treat the classification error as the more serious error, and hence focus mainly on that, trying to minimize the errors on the user part on the go. \par

Thus we have our entire space of movements for the letter gestures to be $4^\mathbb{N}$, and, we want to minimize the maximum number of movements made. We need to accommodate 26 letters, hence $n\geq \left\lceil \frac{1}{2}\log_226\right\rceil=3$ would have been sufficient to accommodate all the letters. However, although mathematically justified, for a dextral person, it is not easy to make all movements with equal flexibility. For instance, it is difficult to bend the phone towards the right as flexibly as one moves it to the left. Also, for the metric we use, we do not want the subsequences to match a lot. For example,  consider LUR and LDR. Both of these have same length, and have the first and the last movement common, thus creating a lot of similarity. This reduces our degrees of freedom.\par 

\subsection{Separating the letters}

\subsubsection{The Markings}
When we are writing a word, we begin a sudden jerk in any direction with the phone, and then make our gesture with the phone. With the end of the gesture, we again make a jerk in any direction, marking the end of one letter and the beginning of the next. Here, we have devised an offline method to identify the jerks.
    
\subsubsection{The Analysis Plan}

Note that a jerk, as is defined in the standard physics literature, is the rate of change of acceleration. Hence, to identify jerks, we look into the variation in the acceleration. \par 
    
Hence, we fix a window length, say length 10, and take moving variance. The moving variance at point $i$ is 
\begin{align} 
v_i&=\dfrac{1}{10-1}\sum_{j=i}^{i+9}\left(a_j-\overline{a}_j\right)^2\hspace{1cm}\forall \ i=1(1)(n-9)
\end{align}
where $a_j$ is the resultant acceleration in all the three axes at the $j$th time point, obtained the sum of squares of acceleration in all the axes, and 
\begin{align}
\overline{a}_j&=\dfrac{1}{10}\sum_{k=j}^{j+9}a_k
\end{align}

Thus we use the following idea: We begin shifting the moving variance window from left to right. Whenever we see the moving variance goes above a certain cutoff (which we shall be describing later the process to obtain), we interpret it as a jerk starting. Then again, when the moving variance goes down that cutoff, the jerk has ended. Thus we cut at these two points. This marks the beginning of a letter. We continue looking at the moving variance and again when it rises over a cutoff, we mark the beginning of a jerk and continue as before. \par

One may consider the following analogy: Consider a ship moving along a straight line in the sea, which has a some turbulent pockets on its path. The ship has a turbulence tracking mechanism, which makes the ship go on high alert when a turbulence is registered. It stays on high alert, and when finally it gets out of the turbulent pocket, it's alarm system goes off. The ship is our moving window, and the cutoff point is our alarm mechanism, registering the time of the jerks.\par 

But note that the jerk is detected by the moving variance in its entirety when the entire window gets within the jerk. For this, we require the window length to be at most the jerk length. A window length of 10 is quite satisfactory, as we can see from empirical evidence. Further, a fixed cutoff value could be used if we want the program to be user specific, but it could be fatally detrimental if the user base has much variation. Hence we need an adaptive procedure to determine the cutoff values. 

\subsubsection{Obtaining cutoff: The k-means approach}
To obtain the cutoff, we first cluster our moving variance values into two groups: \textit{high} and \textit{low}. By the very nature of the problem, our data points to be clustered, are the moving variances of the total resultant acceleration, which come in two categories, one for the letter gestures, the lower class (since the variance in acceleration is very low), and one for the jerks, the correspondingly higher class.

We begin with the K-means approach\cite[Chapter~9]{Kmeans&EM-GMM}, with $k=2$.  The cutoff thus obtained is the mean of the lowermost point of the upper cluster and the uppermost point of the lower cluster.

\subsubsection{Obtaining cutoff: The EM-GMM approach}

From the plots, we see that the lower cluster is spread over a very small region as compared to the higher cluster. Thus, due to difference in cluster size,  EM-GMM, i.e., the Expectation Maximization under Gaussian Mixture Model \cite[Chapter~9]{Kmeans&EM-GMM}  approach has an upper hand over the k-means approach. Figure ~\ref{clus1} shows a case where k-means fails to identify a jerk as it cannot encompass the idea of different variation. EM-GMM, on the other hand, tackles this problem efficiently. The cutoffs are obtained by the mean of the lowest point of the upper cluster and the highest point of the lower cluster. Used the R package ClusterR \cite{ClusterR} for this.

\subsubsection{Obtaining Cutoff: The Problems of the Two Methods and Their Solutions- Bagging and Neighbours Together}

Recall that EM-GMM works under the assumption that the data points are independently generated. But our points for classification is the moving variance, which is, by its very nature, not independent, since each data points are the sum of the total acceleration of 10 points, and consecutive points have 9 points in common. In fact the structure of their dependence is positive, i.e, it is more likely that the next point belongs to the same cluster as this point. \par 

\begin{figure*}
    \centering
    \includegraphics[width=\textwidth]{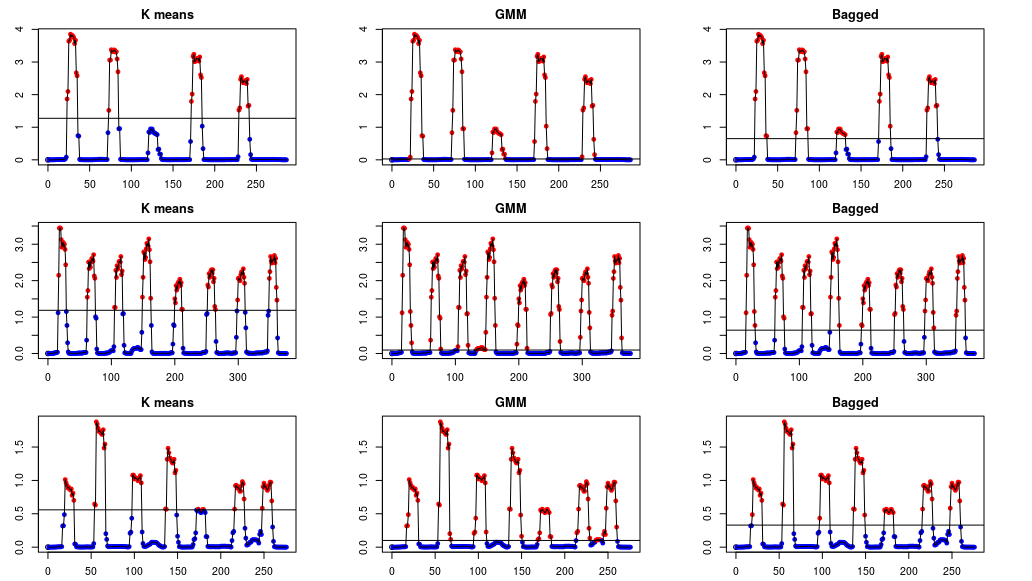}
    \caption{Time VS Moving Variance plot; (top) K-means fails; (middle) GMM fails; (bottom) Both fails. In all the three cases, bagged performs correctly}
    \label{clus1}
    
\end{figure*}

Both the errors can be very fatal, one would completely ignore a jerk and combine two gestures with random noise in between, and give rubbish letters; while the other makes an alphabet into a jerk, thus destroying both the letter and the next letter, as in Figure ~\ref{clus1}

To bring in the advantages of both the k-means approach as well the EM-GMM approach, we use the idea of Bagging predictors \cite{Bagging}: take the cutoff to be the average of the cutoffs given by both the methods. This does not give a very far cutoff so as to miss jerks, but again does not give very conservative cutoff so as to almost brush against the lower class. Figure ~\ref{clus1} shows a situation where both k-means and EM-GMM fail, but bagging helps.\par 

Also, we define an idea called \textbf{Neighbours together}. The idea is as if the points have a tendency to keep their neighbours together. Hence it is not easy for a point to rise to the above cluster if its previous immediate neighbours are in the lower cluster. Thus we allow a point whose immediate previous neighbour is in the lower cluster, to rise above to the next cluster if and only if both its immediate two next neighbours are in the upper cluster as per the bagged cutoff line. The falling down to the lower cluster is handled analogously by its next neighbours. \par 

This is important as the points are not independent. It can never be intended that a point is in the upper cluster and its immediate neighbours are in the lower cluster, pertaining to a point jerk between two letters. This is tackled by Neighbours Together. This encompasses the sense of association that EM-GMM failed to encounter. Also, it makes the bagged cutoff line not a rigid cutoff, adapting to the situation if necessary.\par

\begin{figure}
    \centering
    \includegraphics[width=0.5\textwidth]{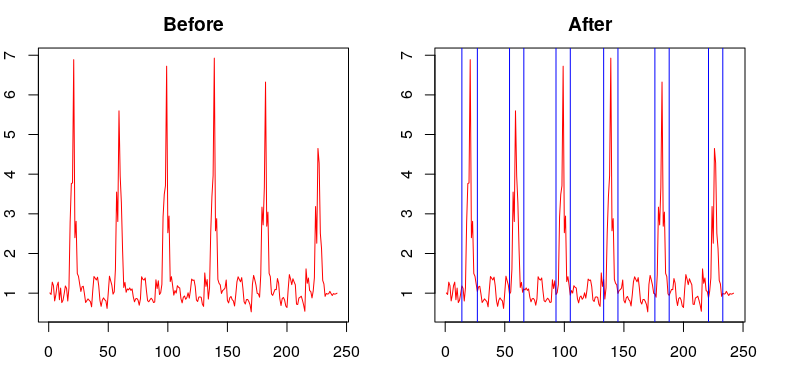}
    \caption{Time Vs squared resultant acceleration plot, before and after obtaining the cutoffs}
    \label{cuts}
\end{figure}

\subsection{Extracting the letters: Scaling and Spline Smoothing}

We define the acceleration points so obtained between the jerks as our letters, and aim to capture the pattern between the letters. For that we need to fit a smooth curve through the points. But this curve should not be an overfit, but  should capture the overall shape of the gesture so created. Hence we use a \textbf{cubic spline smoothing} curve \cite{spsmooth1} \cite{spsmooth2}. \par 
We scale down the curve to a \textbf{unit square} (to tackle variation from the users' end on how much they vary user to user on creating the letters; thus keeping user error in check) and extract it’s values at  100  points.  The  values  at  these  100  points  shall  be  the basis  for  our  comparison  within  the  training  set. We have used a spar value of 0.5 for an optimal tradeoff between overfit and underfit, and the penalizing parameter $\lambda$ has been correspondingly calculated in R. 

\subsection{Using Directionality of Gestures to Reduce confusion: Principal Component Analysis}

\begin{figure}
    \centering
    \includegraphics[width=0.5\textwidth]{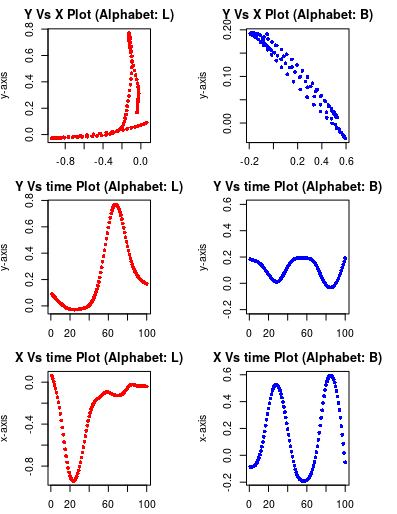}
    \caption{XY plot and across time patters for letters B and L}
    \label{PCAl}
\end{figure}

Table ~\ref{LetterGI} Next, we divide the letters into three categories:  the X axis, the Y axis or the both axis. Figure ~\ref{PCAl}(right) shows the letter B's plot (gesture RR), X versus Y. Note that the X values is from 0 to almost 1, but the Y values has a much lower range. The pattern that is captured is mainly in the X axis, but a component of the pattern is also captured into the Y axis. Thus the principal component of variation is essentially slant towards the X axis. To capture the entire pattern of the data, we rotate the frame of reference into the principal component direction, and then compare the values in the principal direction with the values in the X axis training set. Similarly for the Y axis.\par 

Figure ~\ref{PCAl}(left) shows the letter L's plot (gesture LD), X versus Y. Note that the variation cannot be adequately explained by any one axis only. There is significant pattern in both the X axis and the Y axis, and hence we shall be comparing with the both axis training set. Thus, we perform \textbf{Principal Component Analysis} to decide on the principal axis, and note it's proportion of variation explained. Note that this proportion of variation shall necessarily be at least 50\%. From empirical evidence, we see that the both axes proportion of variance explained by principal component is in the range 75\% to 90\% (since there are gestures where there are two movement in the X axis and one movement in the Y axis, thus it's principal component would explain quite a proportion, but not it's entirety.) But for the truly single axes, the proportion of variance explained is almost always in the range $(97\%,99\%)$. Hence we decide upon the cutoff to be 92\%. Thus, if the proportion of variance is explained by the principal component is $\leq 92\%$, it is classified as both axis.\par 

Else, the gesture is identified in the single axis category. Now, if the principal component direction has a larger magnitude of the X value than the Y value, then the direction is slant towards the X axis, and thus classified X axis. Else it is classified as Y axis. Having decided upon the single axis of variation, the frame of reference for the letter is rotated along the principal component direction, to make the comparison axis the direction of rotation. Note that the rotation is not unique: If $\vec{v}$ is a principal direction, then so is $\vec{v}$. Hence we agree upon the convention that the rotation should be made such that if it is a X axis letter, then it's principal component direction is taken to be positive. Similarly for the Y axis letter. This has a twofold advantage, one to restrict the misclassification to finer subsets of the English alphabet, and the other to make the classification more efficient, capturing the most pattern the data has to offer, even if the user could not provide it in the mentioned axis. 

\subsection{Visualizing the gestures from the Plots}

\begin{figure}
    \centering
    \includegraphics[width=0.5\textwidth]{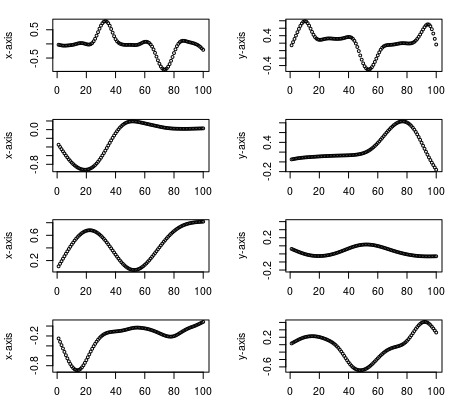}
    \caption{Visualizing Gestures Through plots: Comparing x-axis movement with y-axis movement of various letters. (from top to bottom) Q,L,B,K}
    \label{QLBK}
\end{figure}

Before getting into the details of the metric for comparison, we would like to demonstrate how consistent are gestures are. Refer to Figure ~\ref{QLBK}. Note the plot of B. Recall that the gesture for B is \textbf{RR}. Thus the entire axis of movement is the X axis, as can be seen the corresponding X axis movement, while the Y axis almost remains still. For each right movement, there are undulations in the positive direction when X is plotted against time. There are two undulations clearly visible, from which the letter is clearly visible. Similarly for \textbf{L}, there is a first movement in the X axis towards the down direction, implying a left direction (note that it is opposite direction of right and is thus consistent), when the Y axis remains still, and then there is a movement in the Y direction towards the up, denoting the down movement, when the X axis remains still. Thus the similarity of the curves shows the consistency of the patterns, and gives us the idea that it would provide an efficient mechanism for classification. \par 

Also from the plots, you can guess what the letter is, if the letter is not known, or even guess the gestures. As a fun exercise, the reader may try to recall the gesture of Q from the plots without referring to Table ~\ref{LetterGI}. \par

\subsection{The Metric for Comparison: Soap Bubble Metric}

For the comparison of the test data with the training data, we need an appropriate metric for comparison. For this we define our soap bubble metric. \par 

\begin{figure}
    \centering
    \includegraphics[height=0.4\textwidth,width=0.4\textwidth]{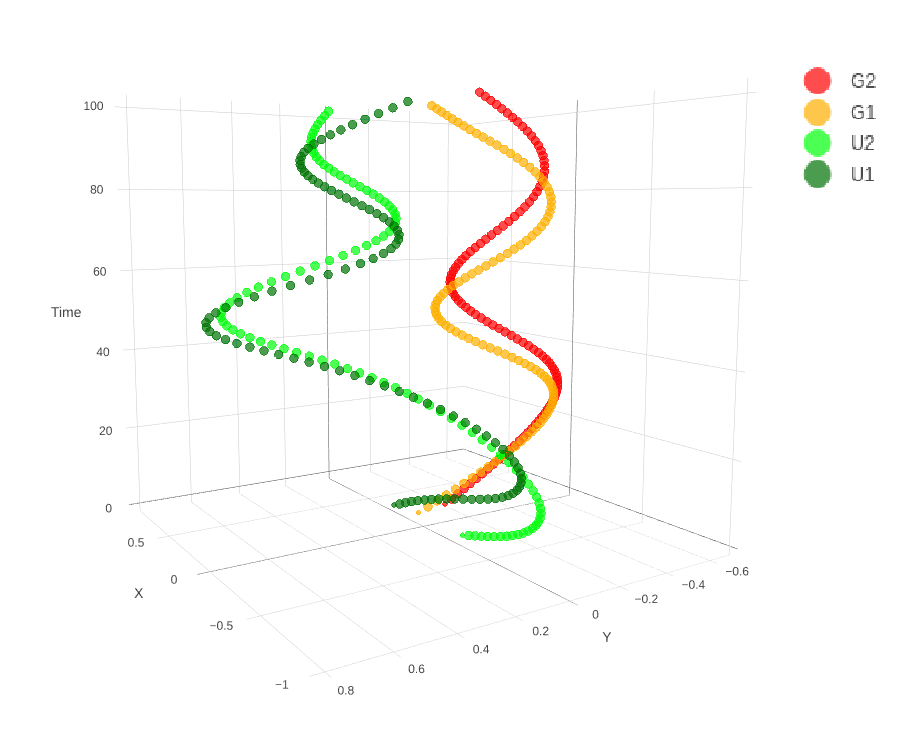}
    \includegraphics[height=0.4\textwidth,width=0.4\textwidth]{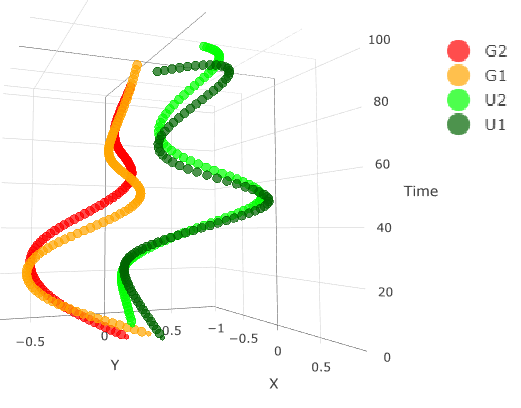}
    \caption{Visualizing the gestures in 3D plane: Both axes- G1 and G2 are two realisations of the letter G, U1 and U2 for U. Notice the proximity of the two G's and U's, while being well separated from each other}
    \label{UVsG}
\end{figure}

First we shall motivate the reader for the soap bubble metric. Suppose you have two inflexible wires of finite length in the form of curves, (Figure ~\ref{UVsG} might help to visualize) and you dip the wires in a soap water solution. Thus a soap film is formed between the wires due to surface tension, and our distance is the area of the soap bubble so formed. Note that for the proper visualizations of the soap film to be formed, we need to have the curves non-orthogonal. To formally define the metric, let $E(\cdot,\cdot)$ be the natural Euclidean metric between two points in the space. Now suppose $\mathbf{a}$ and $\mathbf{b}$ be two curves in the 3D plane, being functions of $t$. The soap bubble metric is defined 
\begin{align} 
d(\mathbf{a},\mathbf{b})=\int_0^n E(\mathbf{a}(t),\mathbf{b}(t))\,dt\approx \sum_{i=1}^n E(\mathbf{a}(t_i),\mathbf{b}(t_i))
\end{align}
where $(0,n]$ is the time interval in which the data is collected, and it is approximated by the distance at the time points at which we collect the data.\par 

\begin{figure}
    \centering
    \includegraphics[width=0.5\textwidth]{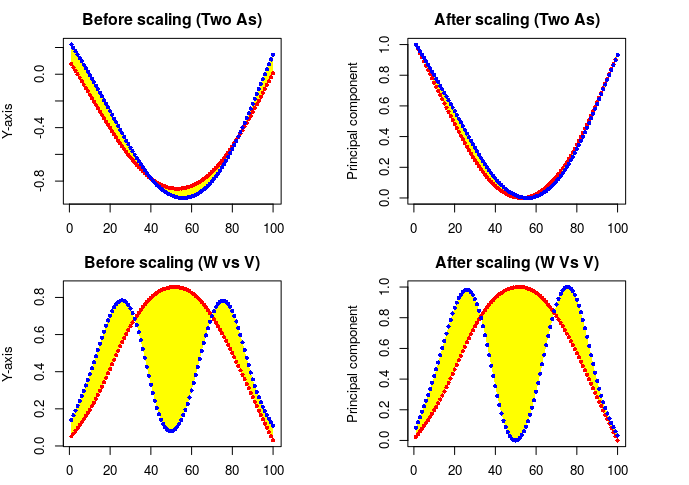}
    \caption{Visualizing the gestures and the areas: Single Axes}
    \label{Single}
\end{figure}

We have added the proof of $d$ being a metric to Appendix \ref{soap}. Note that, for two dimensional curves, say $y(t)$ only, the soap bubble metric essentially boils down to the area between two curves, as shown in Figure ~\ref{Single}

\subsection{The Final Classification}

Having classified the letter as X axis, Y axis or both axis, we only compare distances (the above discussed metric) with the corresponding family. For singles axes, the distance comparison is made only on the single axis, while for both axes, we compare using both the axes. We compute the distances from all the letters in the training set, and classify according to the nearest neighbour. \par 

We had also tried using the 3-Nearest Neighbour Approach, but the updation showed no significant improvement or deterioration. Hence we stick to our initial approach of Nearest Neighbour. \par 

\begin{figure}
    \centering
    \includegraphics[width=0.4\textwidth]{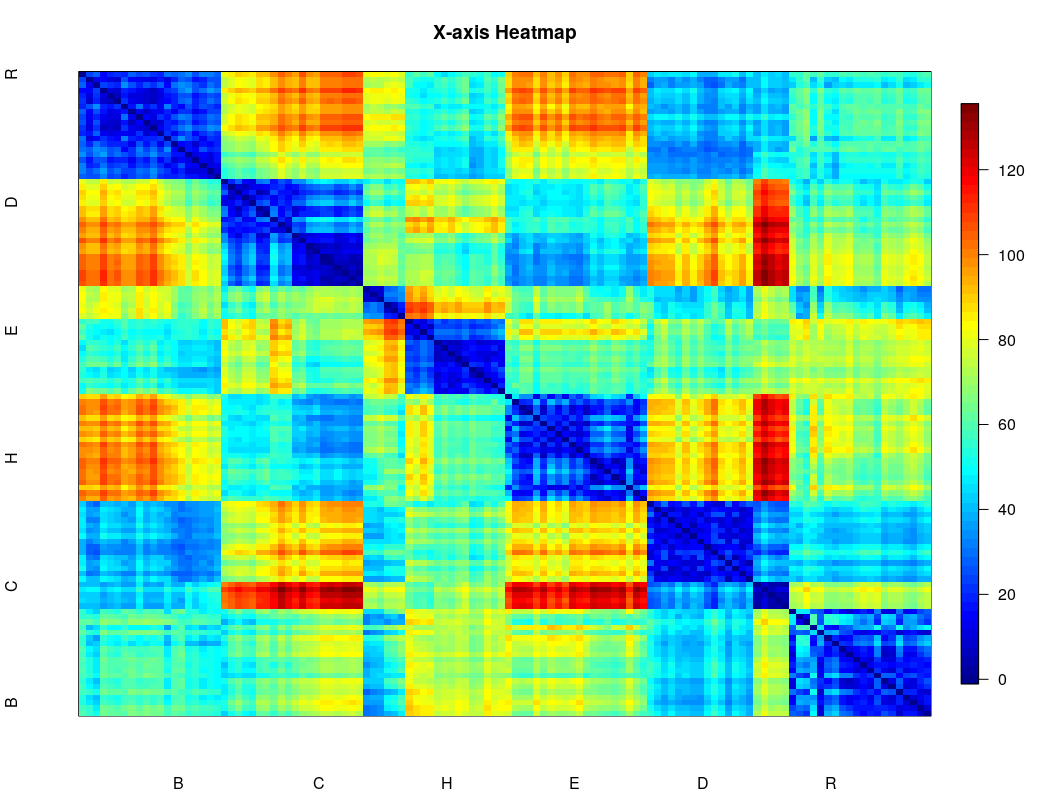}
    \includegraphics[width=0.4\textwidth]{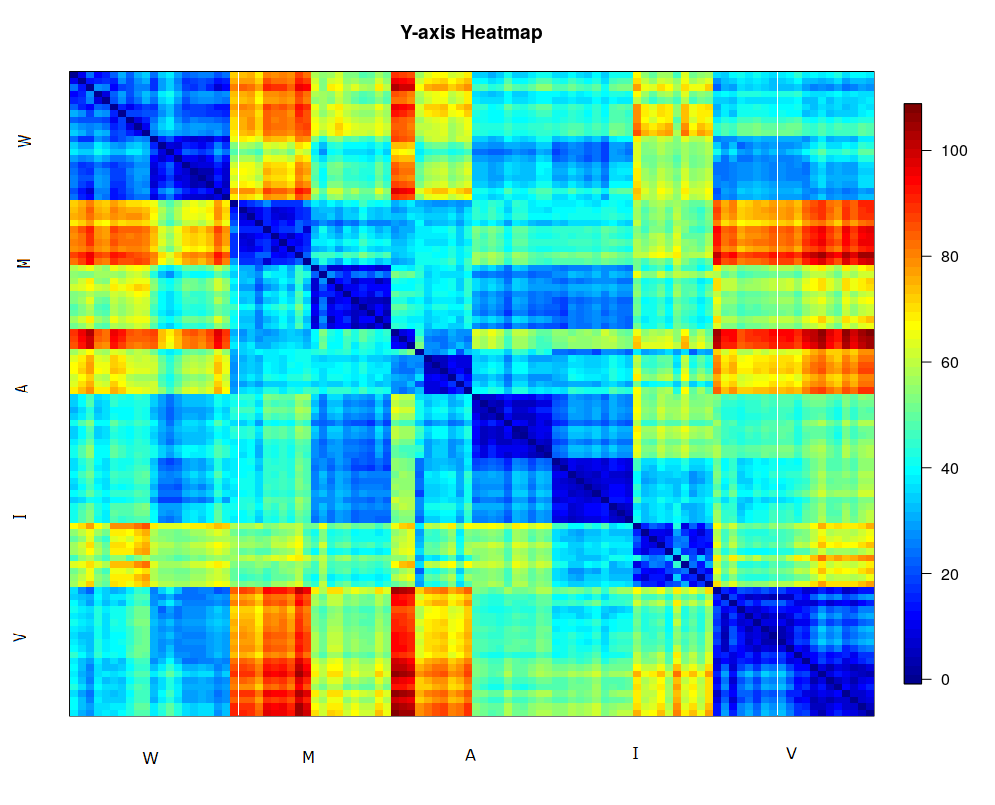}
    \includegraphics[width=0.4\textwidth]{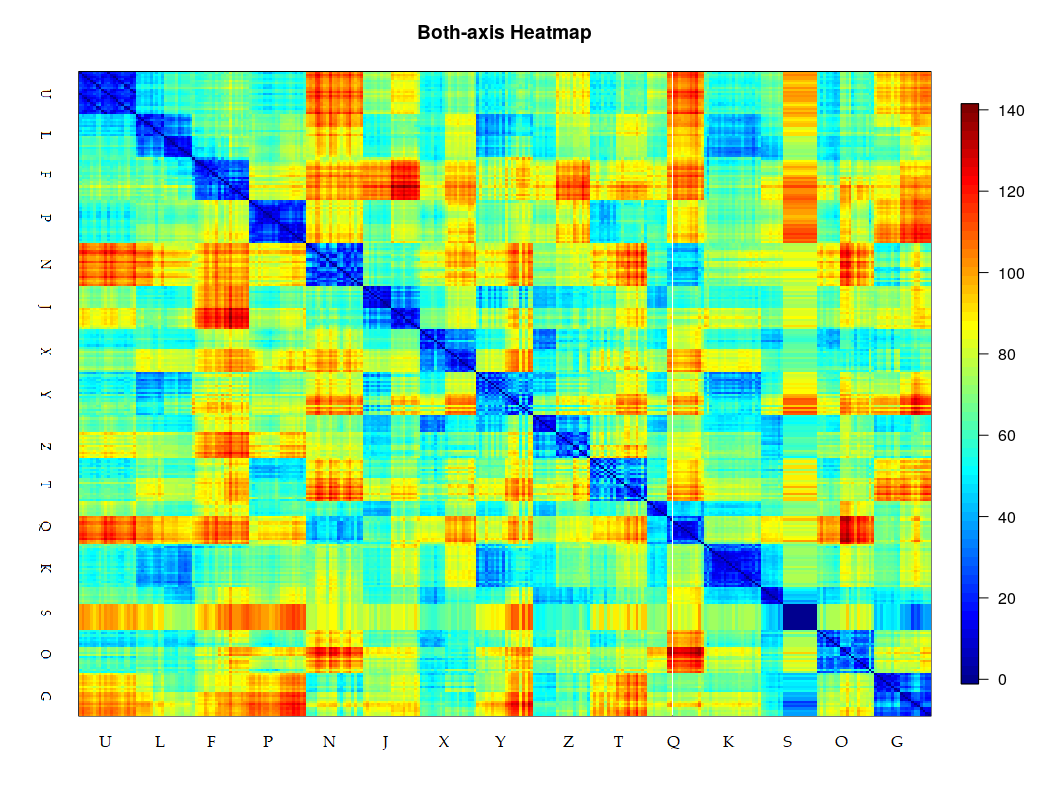}
    \caption{Heatmaps of the corresponding axis letters}
    \label{Heat}
\end{figure}

The heat-maps for the X axis letters, Y-axis letters and both axis letters can be seen in Figure ~\ref{Heat} respectively. The blue end of the spectrum is the least dissimilarity, while the red end represents the most dissimilarity.

\parskip 1pt 

\section{Efficiency of  The Gravity Ink Approach}
\subsection{Expected number of gestures}

Consider a randomly selected text in English, of length $L$ , and we assume that it's letter frequency resembles the letter probability distribution as provided in Table ~\ref{LetterGI}. Thus the expected number of gestures is given 
\begin{align}
    \mathbb{E}(\text{No. of gestures}) &=\left(\underbrace{2.332}_{\text{per letter}}+\underbrace{1}_{\text{jerk}}\right)L+\underbrace{1}_{\text{beginning jerk}}\\ &=\boxed{3.332L+1}
\end{align}


\tikzstyle{decision} = [diamond, draw, fill=blue!20, 
    text width=5em, text badly centered, node distance=3cm, inner sep=0pt]
\tikzstyle{block} = [rectangle, draw, fill=blue!20, 
    text width=5em, text centered, rounded corners, minimum height=4em]
\tikzstyle{block2} = [rectangle, draw, fill=blue!20, 
    text width=8em, text centered, rounded corners, minimum height=4em]
\tikzstyle{line} = [draw, -latex']
\tikzstyle{cloud} = [draw, ellipse,fill=red!20, node distance=5cm, text centered,
    minimum height=2em, text width = 1.6cm]

\begin{figure}[ht]
\begin{tikzpicture}[trim left=-1.15cm]

    \node [cloud] (motion){Motion of phone} ;
    \node [block2, right=1cm of motion] (accelaration){Calculate Resultant Acceleration} ;
    \node [block2, below = 0.5cm of accelaration] (movVAR) {Calculate Moving Variance};
    
    \node [block, below right=0.5cm and 0.2cm of movVAR] (Kmeans) {K-means Clustering (k=2)};
    \node [block, below left=0.5cm and 0.2cm of movVAR] (GMM) {EM-GMM (k=2)};
    
    \node [block, below left=0.5cm and 0.2cm of movVAR] (GMM) {EM-GMM (k=2)};
    
    \node [block2, below =2.5cm of movVAR] (cutoff) {Obtaining Cutoff using Bagging and keeping neighbours together};
   
    \node [block2, below =0.5cm of cutoff] (extracting) {Extract letters and spline smoothing};
    
    \node [decision, below =0.5cm of extracting] (PCA) {Detect axis using PCA};
    
    \node [cloud, left =1cm of PCA] (letters) {For each letter};
    
        \node [block, below =0.5cm of PCA] (Y) {Project on  Y-axis and scale the data};
    \node [block, left =1cm of Y] (X) {Project on  X-axis and scale the data};
    \node [block, right =1cm of Y] (both) {Scale the data};
    
    \node [block, below =0.5cm of Y] (Y2) {Search among Y-axis letters};
    \node [block, below =0.5cm of X] (X2) {Search among X-axis letters};
    \node [block, right =1cm of Y2] (both2) {Search among both axis letters};
    

     \path [line] (motion) -- (accelaration);
     \path [line] (accelaration) -- (movVAR);
     \path [line] (movVAR) -- (GMM);
     \path [line] (movVAR) -- (Kmeans);
     \path [line] (GMM) -- (cutoff);
     \path [line] (Kmeans) -- (cutoff);
     \path [line] (cutoff) -- (extracting);
     
     \path [line] (letters) -- (PCA);
     
     \path [line] (PCA) -- node[right]{if Y-axis}(Y);
     \path [line] (PCA) -- node[left]{if X-axis}(X);
     \path [line] (PCA) -- node[right]{if both-axis}(both);
     
     \path [line] (X) -- (X2);
     \path [line] (Y) -- (Y2);
     \path [line] (both) -- (both2);

\end{tikzpicture}
\label{flowchart}
\caption{Flowchart for Gravity Ink Algorithm}
\end{figure}
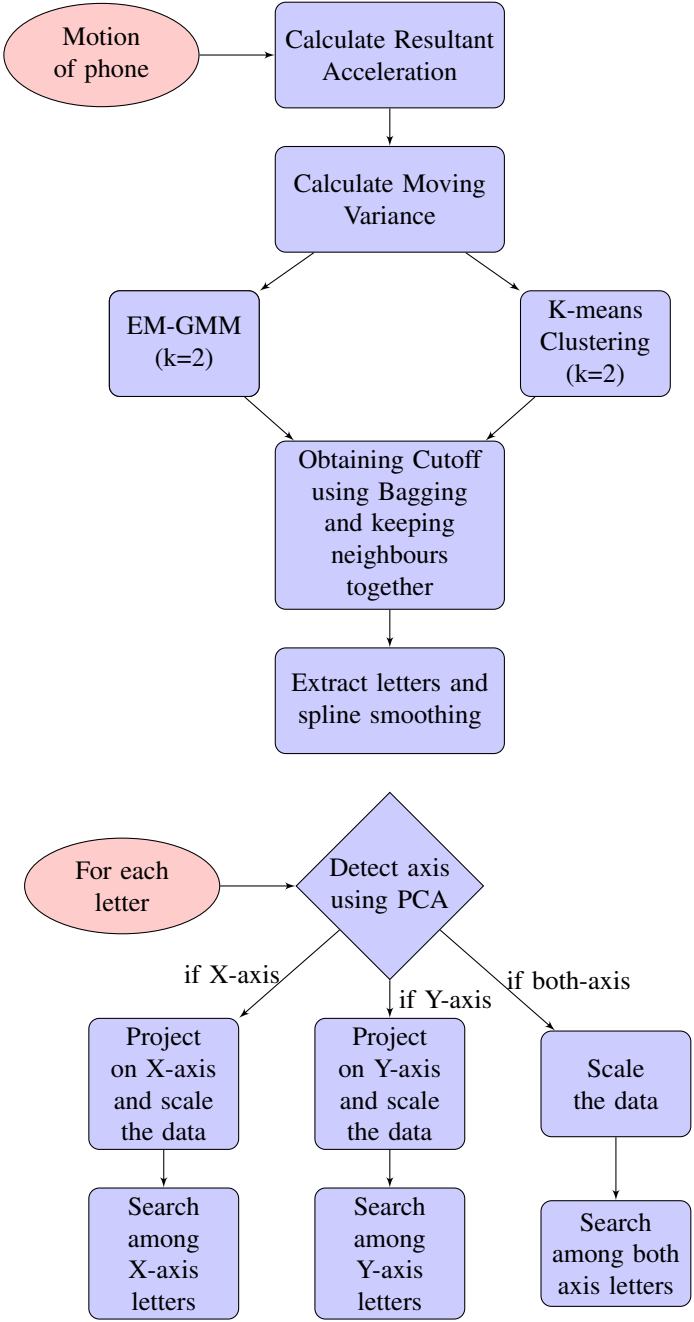

\subsection{Time Complexity}

 Suppose $T$ is the number of training samples, $m$ is the number of letters generated, and $n$ is the total number of data points. 
\begin{itemize}
    \item \textbf{Calculate total acceleration at each point}: $O(n)$
    \item \textbf{k-means algorithm (k=2)}: Distance from the 2-means calculated and means updated iteratively. Hence time complexity: $O(n^2)$
    \item \textbf{EM-GMM}: Gaussian clusters formed in $i$ iterations. No theoretical bounds on $i$ found yet. Hence complexity $O(ni)$ since likelihood is computed for each of the letters.R bounds the maximum number of iterations.
    \item \textbf{Final cutoff using bagging and Nearest Neighbour}: $O(n)$ since Neighbours Together requires $O(n)$ time to scan through all the points.
    \item \textbf{Spline Smoothing}: This is the most dominant term of our algorithm. In one dimension, it is dominated mainly by a matrix inversion, which can be computed in $O(n^{2.373})$ time, via, Optimized Coppersmith– Winograd algorithm.\cite{omatinv} The algorithm used is Reinsch's algorithm for spline smoothing. 
    \item Principal Component Analysis is always computed at 100 points for the two co-ordinates. Thus it is an $O(1)$ operation for each letter, overall $O(m)$
    \item Distance from all elements of Training set to compute nearest neighour: $O(T)$ for each letter. Overall $O(mT)$
\end{itemize}

Hence overall complexity turns out to be \begin{align} \mathbb{O}(n^{2.373}+ni+mT) \end{align}

\section{Simulations: Error Bounds on The Gravity Ink Approach}

To have an idea of how well our Gravity Ink Approach works, we tried to estimate and provide a confidence interval for the error probabilities. Assume, that for each letter, the error probabilities remain unchanged. \par 

Suppose we generate $k$ random letters from the English alphabet, with the letter frequencies as provided in Table ~\ref{LetterGI}. Then, for each letter so obtained, we generate n samples by motion of the phone, typed continuously. Note that, in this section, we concentrate on the error probability $p$ which is essentially the classification error, as it shall remain unchanged over time. The user error is not a fixed probability and it reduces over time, with practice and muscle memory. Hence, instead of generating words, we generate letters and create a continuous stream of n repeated same letters. Thus the classification error is only captured here, as repeatedly creating the same letter almost yields the user error insignificant. \par 

Define 
\begin{align}
    \gamma& =\sum_{i=1}^k\sum_{j=1}^{26} X_i\mathbf{1}_{\{a_i=A_j\}}
\end{align}

where $X_i$ is the number of errors when the $i$-th letter is being created, $i=1(1)k$, and $\mathbf{1}_{\{a_i=A_j\}}$ is the indicator of the event that the $i$th letter created, $a_i$, is the letter $A_j$ of the English alphabet, $j=1(1)26$, where we correspond $A_1$ to A, $A_2$ to B, $A_3$ to C and so on. 
This, after some cumbersome algebraic calculations (which has been moved to Appendix \ref{error bound}) leads us to a $100(1-\alpha)\%$ confidence interval of $p$ to be \begin{align} \left(\widehat{p}-z_{\frac\alpha 2}\sqrt{\dfrac{\widehat{p}(1-\widehat{p})}{nk}},\widehat{p}+z_{\frac\alpha 2}\sqrt{\dfrac{\widehat{p}(1-\widehat{p})}{nk}}\right) \end{align} where $\widehat{p}$, the point estimate of $p$, is given by \begin{align} \widehat{p}=\dfrac{\gamma}{nk} \end{align}

We shall use this formulae to obtain our empirical results. \vspace{0.1cm} \\

$\bullet$ \textbf{Results}: 

Our training set consists of $20$ realisations of each letter, generated and labelled manually by us. Each of them have been smoothed and scaled as required, from beforehand. In order to estimate $p$, we had a test subject who did not contribute to our training set. We randomly chose $k=30$ letters, with replacement, from the English alphabet. He was then asked to type all of these letters, $n=5$ times each and the miss-classifications ($\gamma$) were counted. \par

\vspace{0.25cm}

We had taken $n=5$, $k=30$. The obtained realisation of $\gamma$ was $4$. Hence point estimate for $p$ is \begin{align} \widehat{p}=\dfrac{4}{150}=0.026 \end{align} and a 95\% Confidence Interval is given by \begin{align} (0.0009,0.0524)\ni p\text{ wp } 0.95 \end{align}

\section{Conclusion and Further Scope}

In this paper we have come up with a smartphone-accelerometer based gesture typing method, which has been designed to make the gestures simple yet well separable. The algorithm is simple and intuitive, and has been made so that it does not require many training samples, without compromising on accuracy (like the use of PCA or neighbours together). Moreover, it has the flexibility to be made user specialized, replacing the preset training samples with the user inputs, that would enhance accuracy and hence user satisfaction. \par 

The error bounds we have presented are essentially empiric, and an alternate way could be finding out the theoretical probability of errors. Moreover, the PCA cutoff chosen here is essentially subjective, and it can be improved upon- a possible way could be using the Wald's sequential approach. The idea being that we take random samples from the training set of one axis and from the training set of both axes, and if the test sample proportion of variance is explained is close enough, then it is classified accordingly, else further samples are taken iteratively. However then, the convergence and error bounds need to be sufficiently taken care of. \par 

Our offine method of variable separation can be replaced by some corresponding online version, when, the letters can be separated on the go. \par

The Nearest Neighbour approach could be altered by a more sophisticated kernel based approach, using a mean curve for the movement, and the deviations a white noise. The almost unused Z-axis can be used to introduce spaces, backspaces and punctuation into the alphabet. A mobile app implementing this algorithm could then be thought to be realised. 


\appendices
\section{Proof of Soap Bubble Metric being a metric}{\label{soap}}

$$d(\mathbf{a},\mathbf{b})=\int_0^n E(\mathbf{a}(t),\mathbf{b}(t))\,dt\approx \sum_{i=1}^n E(\mathbf{a}(t_i),\mathbf{b}(t_i))$$ where $E(\cdot,\cdot)$ stands for the standard Euclidean metric between two points.

\begin{itemize}
    \item $d(\mathbf{a},\mathbf{b})\geq 0$: The Euclidean distance is itself non-negative, hence it's integral is also positive, unless the curves are identical throughout.
    \item \textbf{Equality iff curves are same}: If condition is trivial. For the only if condition, note that $E(\cdot,\cdot)$ is a non-negative function, hence it's integral is 0 implies the $E(\cdot,\cdot)$ is zero, assuming the curves are continuous functions of time. Thus the iff condition holds.
    \item \textbf{Commutativity}: Clearly $d(\cdot,\cdot)$ is commutative, because $E(\cdot,\cdot)$ is commutative.
    \item \textbf{Triangle Inequality}: Note that, by the inherent property of surface tension, it tends to create the lowest energy minimizing position, which is, in our case, is the minimum area. Thus, suppose three curves, $\mathbf{f}$, $\mathbf{g}$ and $\mathbf{h}$, we must have $d(\mathbf{f},\mathbf{g})\leq d(\mathbf{f},\mathbf{h})+d(\mathbf{h},\mathbf{g})$ due to surface tension, because, had the inequality been strict in the opposite direction, then due to surface tension, it would have rather formed the soap film via $h$ than directly through $\mathbf{f}$ and $\mathbf{g}$. This is a contradiction, as the surface tension is the area minimizing position.\par 
    To prove it rigorously, note that $E(\cdot,\cdot)$ itself maintains the triangle inequality. Thus \begin{align*} d(\mathbf{f},\mathbf{g})&=\int_0^n E(\mathbf{f}(t),\mathbf{g}(t))\,dt\\ &\geq \int_0^n E(\mathbf{f}(t),\mathbf{h}(t))\,dt +\int_0^n E(\mathbf{g}(t),\mathbf{h}(t))\,dt\\ &= d(\mathbf{f},\mathbf{h})+d(\mathbf{h},\mathbf{g}) \end{align*}
\end{itemize}
This concludes our proof. \hfill $\blacksquare$

\section{Point and interval estimate calculations for Error bounds}{\label{error bound}}

Define 
\begin{align*}
    \gamma& =\sum_{i=1}^k\sum_{j=1}^{26} X_i\mathbf{1}_{\{a_i=A_j\}}
\end{align*}
where $X_i$ is the number of errors when the $i$-th letter is being created, $i=1(1)k$, and $\mathbf{1}_{\{a_i=A_j\}}$ is the indicator of the event that the $i$th letter created, $a_i$, is the letter $A_j$ of the English alphabet, $j=1(1)26$, where we correspond $A_1$ to A, $A_2$ to B, $A_3$ to C and so on. 

Since the samples are independent, hence $X_i$ and $X_j$ are independent for $i\neq j$, and similarly are $a_i$ and $a_j$. Also, $X_i|a_i\sim\text{Bin}(n,p)\forall i=1(1)k$ under the assumption that the samples are created independently and identically. Since the conditional distribution of $X_i$ given $a_i$ does not involve $a_i$ as a parameter, it is easy to argue that $X_i$ and $a_i$ are independent, and the marginal of $X_i$ is the same as its conditional distribution. We denote the probability that $a_i$ assumes the value $A_j$ with probability $p_j$, $j=1(1)26$ for all $i$. Note that $\gamma$ is essentially all the errors formed in the sample. 

Now
\begin{align*}
    \mathbb{E}(\gamma)&=\mathbb{E}\left(\sum_{i=1}^k\sum_{j=1}^{26} X_i\mathbf{1}_{\{a_i=A_j\}}\right)\\
    &=\sum_{i=1}^k\sum_{j=1}^{26}\mathbb{E}\left(X_i\mathbf{1}_{\{a_i=A_j\}}\right)\\
    &=\sum_{i=1}^k\sum_{j=1}^{26}npp_j\\
    &=nkp\cancelto{1}{\left(\sum_{j=1}^{26}p_j\right)}\\
    &=nkp\\
    \therefore \widehat{p}&=\dfrac{\gamma}{nk}
\end{align*}
Hence 
\begin{align*}
    \Var(\gamma)&=\Var\left(\sum_{i=1}^k\sum_{j=1}^{26}X_i\mathbf{1}_{\{a_i=A_j\}}\right)\\
    &=\sum_{i=1}^k \Var\left(\sum_{j=1}^{26}X_i\mathbf{1}_{\{a_i=A_j\}}\right)
\end{align*}

Suppose $X$ and $Y$ are independent random variables, then 
\begin{align*}
    \Var(XY)&=\mathbb{E}(\Var(XY|Y))+\Var(\mathbb{E}(XY|Y))\\
    &=\mathbb{E}(Y^2\Var(X|Y))+\Var(Y\mathbb{E}(X))\\
    &=\mathbb{E}(Y^2\Var(X))+\Var(Y\mathbb{E}(X))\\
    &=\mathbb{E}(Y^2)\Var(X)+\Var(Y)\mathbb{E}(X)^2\\
    &=\Var(X)\mathbb{E}(Y)^2+\Var(Y)\mathbb{E}(X)^2+\Var(X)\Var(Y)
\end{align*}

Here $X_i$ and $a_i$ are independent. Thus we use 
\begin{align*}
    \Var\left(\sum_{j=1}^{26}X_i\mathbf{1}_{\{a_i=A_j\}}\right)
    &=\sum_{j=1}^{26}\Var(X_i\mathbf{1}_{\{a_i=A_j\}})\\ &+\sum_{j\neq j'}\Cov(X_i\mathbf{1}_{\{a_i=A_j\}},\\
    & \hspace{0.1\textwidth}X_i\mathbf{1}_{\{a_i=A_{j'}\}})
\end{align*}

Now, 
\begin{align*}
    \Var(X_i\mathbf{1}_{\{a_i=A_j\}})&=\Var(X_i)\mathbb{E}(\mathbf{1}_{\{a_i=A_j\}})^2\\
    &+\Var(\mathbf{1}_{\{a_i=A_j\}})\mathbb{E}(X_i)^2\\
    &+\Var(X_i)\Var(\mathbf{1}_{\{a_i=A_j\}})\\
    &=np(1-p)p_j^2+(p_j-p_j^2)(np)^2\\ &+np(1-p)(p_j-p_j^2)\\
    \therefore \sum_{j=1}^{26} \Var(X_i\mathbf{1}_{\{a_i=A_j\}})&=-(np)^2\sum_j p_j^2\\ &+[(np)^2+np(1-p)]\cancelto{1}{\sum p_j}
\end{align*}

\begin{align*}
    \Cov(X_i\mathbf{1}_{\{a_i=A_j\}},X_i\mathbf{1}_{\{a_i=A_j'\}})&=\mathbb{E}(X_i^2\mathbf{1}_{\{a_i=A_j\}}\mathbf{1}_{\{a_i=A_{j'}\}})\\&-\mathbb{E}(X_i\mathbf{1}_{\{a_i=A_j\}})\\&\hspace{0.1\textwidth}\mathbb{E}(X_i\mathbf{1}_{\{a_i=A_{j'}\}})\\
    &=0-(np)^2p_jp_{j'}
\end{align*}

Thus 
\begin{align*}
    \Var\left(\sum_{j=1}^{26}X_i\mathbf{1}_{\{a_i=A_j\}}\right)&=-(np)^2\left(\sum_j p_j^2+\sum_{j\neq j'}p_jp_{j'}\right)\\&+(np)^2+np(1-p)\\
    &=-(np)^2\cancelto{1}{\left(\sum_{j=1}^{26}p_j\right)^2}\\ &+(np)^2+np(1-p)\\
    &=np(1-p)
\end{align*}

Hence, $$\Var(\gamma)=nkp(1-p)$$

Since $X_i\mathbf{1}_{\{a_i=A_j\}}$ are all independent and identical units, hence by Central Limit Theorem, we know that $$\dfrac{\left(\gamma-nkp\right)}{\sqrt{nkp(1-p)}}\xrightarrow{\text{d}}N(0,1)$$ and since $\widehat{p}$ is a consistent estimator for $p$ (variance goes to 0, hence Chebyshev's inequality assures consistency), hence by Slutsky's theorem, we finally write $$\dfrac{\sqrt{nk}(\widehat{p}-p)}{\sqrt{\widehat{p}(1-\widehat{p})}}\xrightarrow{\text{d}}N(0,1)$$

 Thus a $100(1-\alpha)\%$ Confidence Interval for $p$ is given by $$\left(\widehat{p}-z_{\frac\alpha 2}\sqrt{\dfrac{\widehat{p}(1-\widehat{p})}{nk}},\widehat{p}+z_{\frac\alpha 2}\sqrt{\dfrac{\widehat{p}(1-\widehat{p})}{nk}}\right)$$


\ifCLASSOPTIONcompsoc
  \section*{Acknowledgments}
\else
  \section*{Acknowledgment}
\fi

The authors would like to thank Dr. Arnab Chakraborty, Applied Statistics Unit, ISI Kolkata, and Prof. Bimal Roy, Applied Statistics Unit, ISI Kolkata; for the introduction to the problem and supervision to our methods. We'd also like to thank the Accelerometer Sensor app, which provides data on the acceleration values in a csv file format. It is freely available on Google Play Store.


\ifCLASSOPTIONcaptionsoff
  \newpage
\fi



\bibliographystyle{IEEEtran}
%

%

\vspace{20Pt}

\begin{IEEEbiography}[{\includegraphics[width=1in,height=1.25in,clip,keepaspectratio]{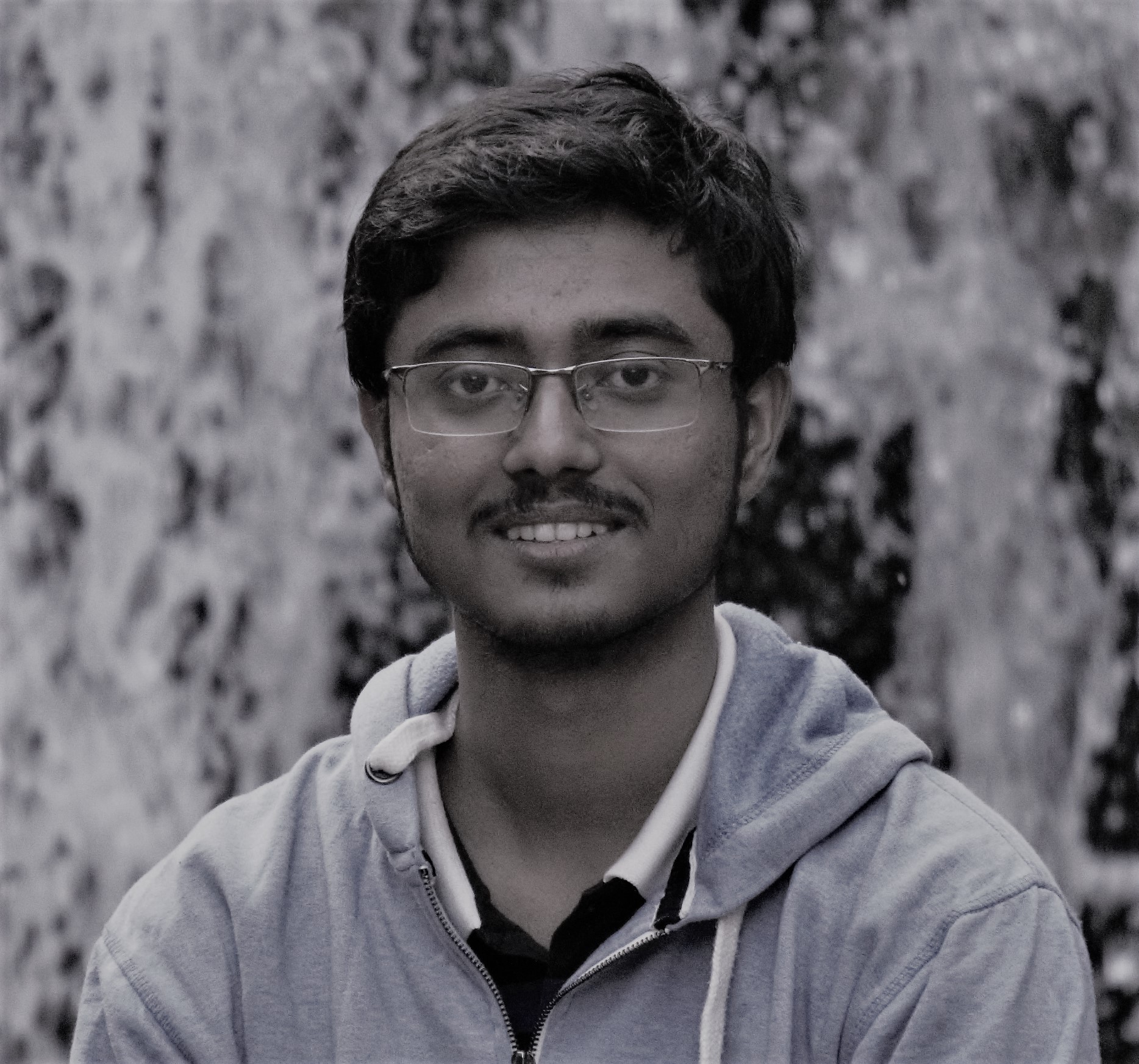}}]{Arindam Roy Chowdhury}

 completed his Bachelor of Statistics with Honours from Indian Statistical Institute, Kolkata and is currently pursuing M.Stat ($1^{st}$ year) from the same. He is a former intern at the Centre of Science for Student Learning (CSSL). His research interests include algorithms, hand gesture recognition, machine learning  and applied statistics. 
\end{IEEEbiography}

\vspace{-30Pt}
\begin{IEEEbiography}[{\includegraphics[width=1in,height=1.25in,clip,keepaspectratio]{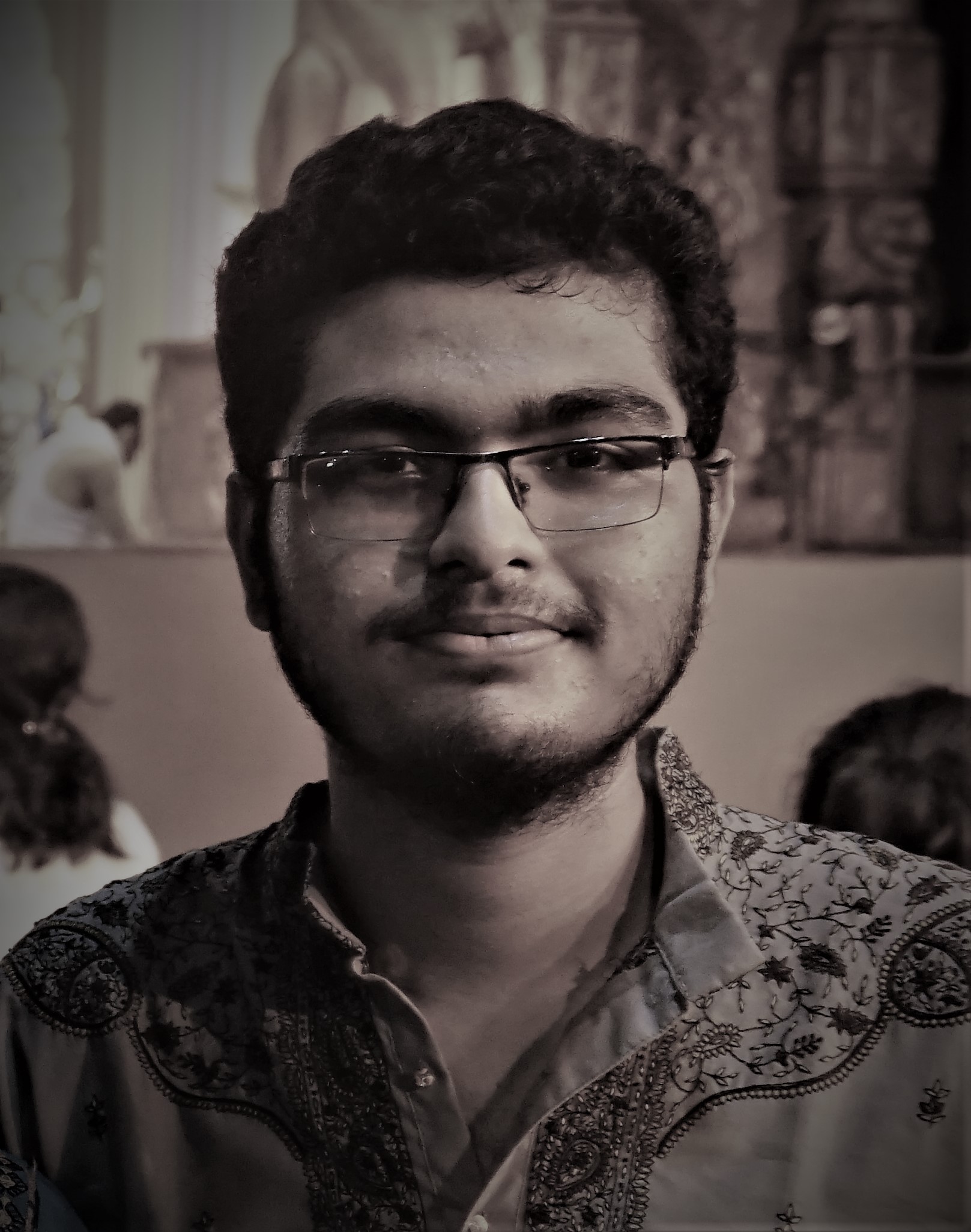}}]{Abhinandan Dalal}
completed  Bachelor of Statistics with Honours  from  Indian  Statistical  Institute  Kolkata, and  currently pursuing Master of Statistics in his First Year as a full  time  student.  He  has  had  an  experience of  an  academic internship  at  DST-  Centre  for Policy  Research,  IISc Bangalore.  His  research interests  include  Statistics  and Economics,  in particular, Hand gesture recognition, Machine Learning, Auctions, Econometrics and Applied Statistics.

\end{IEEEbiography}

\vspace{-30Pt}

\begin{IEEEbiography}[{\includegraphics[width=1in,height=1.25in,clip,keepaspectratio]{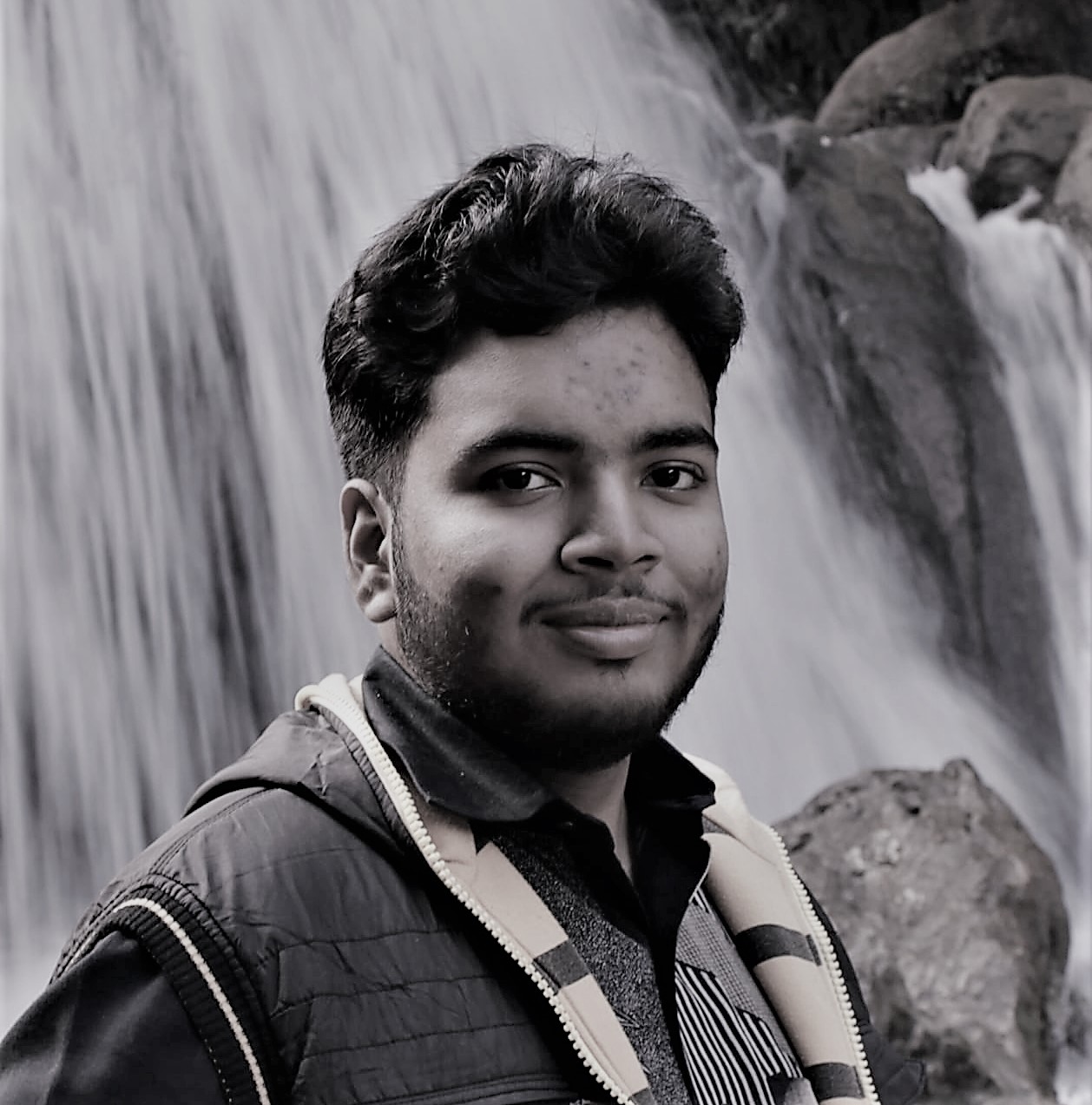}}]{Shubhajit Sen} is a
 Bachelor of Statistics with Honours from Indian Statistical Institute, Kolkata. Currently pursuing Masters of Statistics as a full time student from the same. Has had the experience of a research project at Stowers Institute for Medical research, Kansas City, USA in the summer of 2019. His research interests comprises of hand gesture recognition, Machine Learning, Biostatistics and Applied Statistics.
\end{IEEEbiography}




\end{document}